\documentclass[epjH,final]{svjour}
\usepackage{graphicx}
\usepackage[utf8]{inputenc}
\usepackage{rotating}
\usepackage[table]{xcolor}
\def\degr{\hbox{$^\circ$}}
\def\arcmin{\hbox{$^\prime$}}
\def\arcsec{\hbox{$^{\prime\prime}$}}
\def\utw{\smash{\rlap{\lower5pt\hbox{$\sim$}}}}
\def\udtw{\smash{\rlap{\lower6pt\hbox{$\approx$}}}}

\def\min{\hbox{$^{\rm m}$}}

\def\fs{\hbox{$.\!\!^{\rm s}$}}

\def\farcs{\hbox{$.\!\!^{\prime\prime}$}}

\begin{document}
\title{The first measurement of the deflection of the
vertical in longitude}
\subtitle{The figure of the earth in the early 19th century}
\author{Andreas Schrimpf\thanks{\email{andreas.schrimpf@physik.uni-marburg.de}}}
%
\institute{Philipps-Universität Marburg, Fachbereich Physik, Renthof 5,
D-35032 Marburg, Germany}
\def\titlerunning{A. Schrimpf: The first measurement of the deflection of the
vertical in longitude}
\date{Received 18 December 2013 / Received in final form 20 February 2014 /
Published online xxx}
\abstract{
During the summer of 1837 Christian Ludwig Gerling, a former student of Carl
Friedrich Gauß's, organized the world wide first determination of the deflection
of the vertical in longitude. From a mobile observatory at the Frauenberg near
Marburg (Hesse) he measured the astronomical longitude difference between C.F.\
Gauß's observatory at Göttingen and F.G.B.\ Nicolai's observatory at Mannheim
within an error of 0\farcs4. To achieve this precision he first used a series of
light signals for synchronizing the observatory clocks and, second, he very
carefully corrected for the varying reaction time of the observers. By comparing
these astronomical results with the geodetic--determined longitude differences
he had recently measured for the triangulation of Kurhessen, he was
able to extract a combined value of the deflection of the vertical in longitude
of Göttingen and Mannheim. His results closely agree with modern
vertical deflection data.
} 
\maketitle
\section{Introduction}
\label{intro}
The discussion about the figure of the earth and its determination was
an open question for almost two thousand years, the sciences involved were
geodesy, geography and astronomy. Without precise instruments
the everyday experience suggested a flat, plane world, although ideas of a
spherically shaped earth were known and accepted even in the ancient
world. Assuming that the easily observable daily motion of
the stars is due to the rotation of the earth, the rotational axis can be used
to define a celestial sphere; a coordinate system, where the stars' position is
given by two angles. 
Projecting this celestial sphere on the globe of 
the earth, one can now determine the geographical latitude by observing the
height of stars. The geographical longitude can be deduced from the
meridian transit time of stars. These coordinates are numbers on a perfectly
shaped sphere. By comparing these measurements with those obtained from
field measurements, from a triangulation of the earth's surface,
a more sophisticated model of the figure of the earth appears: the \textit{mean
shape} can be described as an ellipsoid of rotation where, due to centrifugal
forces, the polar diameter is 43 km shorter than the equatorial diameter. The
\textit{real} earth figure, the so called geoid, deviates from the ellipsoid of
rotation in the range of $\pm 100$ m in height. The geoid corresponds to the
equipotential surface of the mean global sea surface, which theoretically will
continue under the continents. The difference between the measured  direction of
gravity and the normal of the ellipsoid of rotation is called the
\textit{deflection of the vertical}. Nowadays it can easily be determined by the
difference of measured stars in zenith direction from the calculated stars using
the ellipsoid of rotation model. The pole flattening of the earth is in the
range of some $10^{-3}$ of the mean diameter, the deviation of the real shape of
the earth from the ellipsoid of rotation is smaller by another three orders of
magnitude. Thus the ability to detect and to measure these deviations reflects
the sensitivity of the instruments and methods available at any particular period.
 
This paper summarizes the historical development of methods to determine the
figure of the earth and then concentrates on the first measurements of a
deflection of the vertical in the first half of the 19th century, especially
the deflection of the vertical in longitude, which is much harder to
observe than that in latitude. It is a masterpiece of an astronomic--geodetic
measurement on the very edge of the possibilities of that time.

\section{The Earth: an ellipsoid of rotation}
\label{sec_ellipsoid_rotation}
Determining of the figure of the earth has been a challenge for
centuries. In the ancient world Eratosthenes (about 240 B.C.) assumed a
spherical shape. He deduced the earth's radius from a
measurement of the zenith angle of the sun at different positions on the earth
and the length of path between those positions, i.e.\ the arc length of the
meridian between the two locations. This is the first known \textit{arc
measurement}: one compares the length of an arc on the earth at a fixed
longitude with the length of the corresponding arc on the sky. Eratosthenes
himself did not actually measure the arc on the earth, but most probably used
the distance between the two positions from the Egypt cadaster maps determined
by step counters \cite{Torge2001}. He achieved a precision of about 10\%.
In the Early Middle Ages Al-Ma'mun, caliph of Baghdad,  commissioned an arc
measurement of 2 degrees and determined the radius of the earth with an error
between 1 and 2\%.

In the 17th century, when the concept of gravity was introduced, scientists
started to ask how measurements of the earth's mean specific weight and
its exact shape could give clues to its internal structure. Using a pendulum
with a period of oscillation proportional to the square root of the ratio of
its length and the acceleration due to gravity, first systematic deviations of the
earth's gravity at different geographic latitudes were found. Isaac Newton
proposed a rotational ellipsoid as an equilibrium figure for a homogeneous fluid
rotating earth with a different curvature at the equator and the poles.
To test this assumption the French Academy of Sciences initiated two arc
measurement campaigns: First, in Peru, at low latitudes, Pierre Bouguer,
Charles de la Condamine and Louis Godin conducted measurements from 1735
to 1744, and second, in 1736/37 Pierre-Louis Moreau de Maupertuis and
Alexis-Claude Clairaut were sent to Lapland for measurements at high latitudes
\cite{Torge2001}. The result of the expeditions' findings was that the diameter
of the earth at the poles is shorter by about 1/300 compared to the diameter at the
equator --- the ellipsoid of rotation as the figure of the earth was born.

The scientists knew that mountains and depths could not be described by a simple
body of rotation. However the value of the earth's rotation is constant within a
precision which could not be achieved in the 18th century, therefor the
measurable mean shape of the earth should be fairly close to a geometrically
defined body of rotation. Commissioned by the French Academy Pierre Méchain 
and Jean-Baptiste
Joseph Delambre organized an arc measurement in France between 1792 and 1798
\cite{Alder2003}.
Equipped with new and more precise instruments their goal was to measure the
precise length of a meridian arc of about 10 degree latitude difference and, by
comparing this with the zenith angle difference of the arc's ends, to determine 
a value for the size of
the earth with a precision not yet attained. After knowing the precise
size of the earth with this value, a new measure of
length was to be defined: the meter.

Overall, the expedition was successful in delivering more precise
measures of the figure of earth. However, it revealed an astonishing
result: the curvature of the meridian arc passing through Paris was larger than
the supposed mean value by a factor of approximately two, leading to a pole
flattening of 1/150 \cite{Laplace1799}. Unfortunately the goal of a generally
accepted definition of the meter could not be achieved. In his final report of
the expedition Delambre combined their results with those of the former Peruvian
arc measurement and finally used 1/334 for determining the length of the meter
\cite{Torge2001}.
Méchain, Delambre and other scientists started to accept that any meridian arc
of the earth features its own curvature and, considering the precision of
measure achievable at the end of 18th century, that the earth
could no longer be described as a symmetric body of rotation. However, in the
following years the main goal was to precisely describe the
\textit{mean} ellipsoid of rotation of the earth. Henrik Johan Walbeck
determined a flattening of 1/302.78 from five arc measurements \cite{Gauss1828}.
This numerical value was used by Carl Friedrich Gauß and Christian Ludwig
Gerling for their triangulations. From the results of ten different
arc measurements and a further correction of the French arc measurement
Friedrich Wilhelm Bessel calculated an oblateness of 1/299.1528
[Bessel 1837;Bessel 1841]%
. As of 1979 the
ellipsoid defined by GRS80 (geodetic reference system 1980) with a pole flattening of 1/298.257222101 is
the recommended value of the best description of a global reference ellipsoid.

\section{Christian Ludwig Gerling}
\label{sec_gerling}
\begin{figure}[h]
\begin{center}
\resizebox{0.75\columnwidth}{!}{\includegraphics{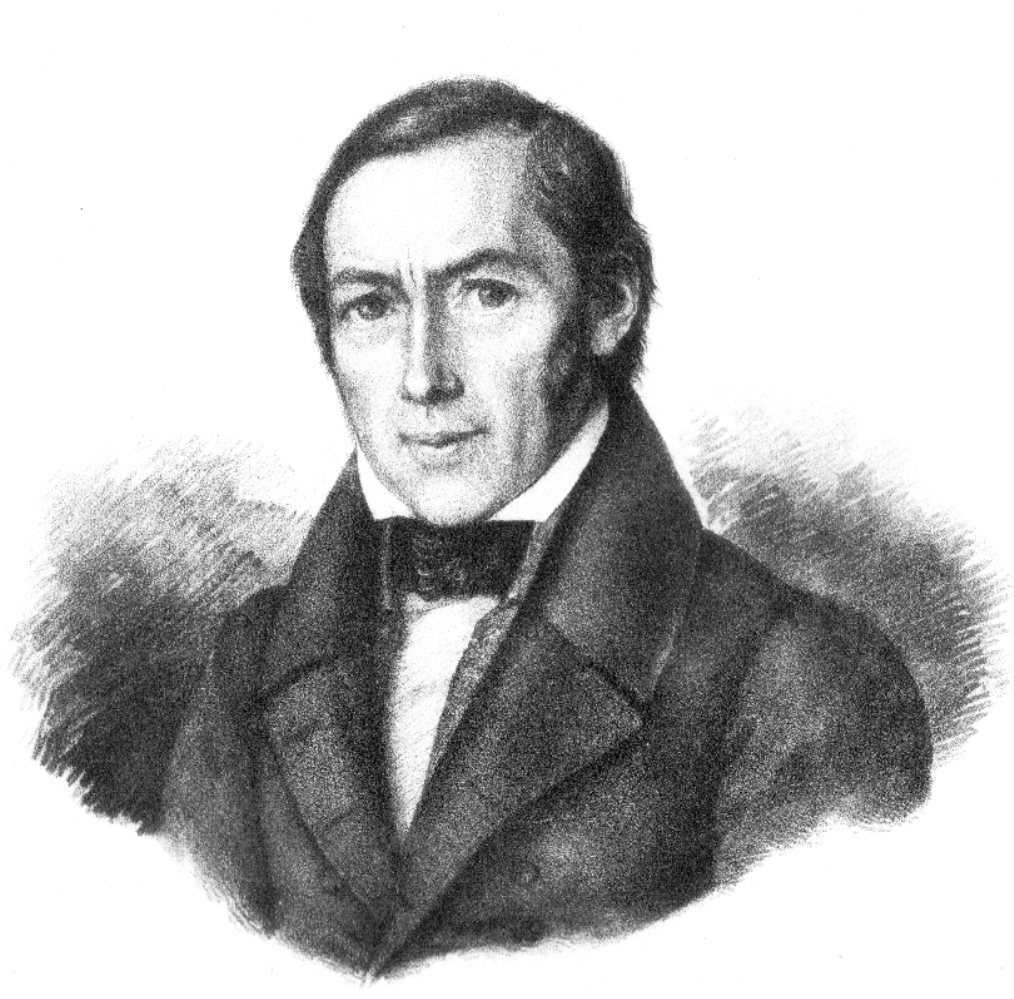}}
\caption{Christian Ludwig Gerling (1788--1864)} 
\label{Fig_Gerling}
\end{center}
\end{figure}

Christian Ludwig Gerling (Fig.\ \ref{Fig_Gerling}) was born in Hamburg, Germany,
in 1788. He was educated together with his longtime friend Johann Franz Encke,
who later became director of the Berlin Observatory. After finishing school,
Gerling attended the small University of Helmstedt, but in 1810 he continued his
academic education in the fields of mathematics, astronomy, physics and
chemistry at the University of Göttingen. He started working at the observatory
of Göttingen under Carl Friedrich Gauß and Karl Ludwig Harding, and, after
some visits in 1811 to the observatories of Gotha (Seeberg), Halle and Leipzig,
he completed his PhD in 1812. 

After he received his PhD Gerling entered a position at a high school in Cassel,
Hesse. At that time he used a small observatory in Cassel for astronomical
observations and occupied himself with calculating the ephemerides of the
asteroid Vesta. He continued to seek a university position and finally in
1817 was appointed full professor of mathematics, physics and astronomy and
director of the ''Mathematisch--Physikalisches Institut'' at the
Philipps--Universität of Marburg. In spite of several offers elsewhere, he
remained at the university in Marburg until his death in 1864 \cite{Madelung1996}.

Gerling's scientific work was affected by two mayor topics. In his early
period in Marburg from 1817 to 1838 he was rather occupied with organizing
the main triangulation of Kurhessen, the data analysis and publication of the
results.

In 1838 the institute moved to a new home in Marburg at the ''Rent\-hof''. After
the building was reconstructed in 1841, he could finally put into operation his
new but small observatory, built on top of a tower of Marburg's old city wall
\cite{Schrimpf2010}. Gerling pursued the scientific topics of astronomy of that
time, making meridional observations and differential extra-meridional
measurements of stars, planets and asteroids, observations of lunar
occultations, etc., mainly to improve the precision of star catalogs and orbit
parameters of solar system bodies.

Carl Friedrich Gauß and Christian Ludwig Gerling's relationship began as a
teacher and student, but during the following years they became each others'
counselor and finally close friends. The correspondence between Gerling and Gauß
not only contains details of scientific discussions but also reflects their
close relationship 
[Schäfer 1927;Gerardy 1964]%
.
Gauß taught Gerling the careful and correct use of scientific
instruments and also the mathematical methods necessary for reducing geodetic
and astronomic measurements. Gerling published a widely used textbook about
planar and spherical trigonometry (the last edition published posthumously in
1865) and he became well known as a teacher in the practical use of many of the
methods Gauß developed theoretically, for example the use of least squares in
geodesy \cite{Gerling1843}.
Gerling died in 1864 at the age of 76. Johann Jacob Baeyer wrote in his
obituary: 
%
%
''\textit{With him, the 'Mitteleuropäische Gradmessung' \em[Central European
Arc Measurement]\em\  has lost a geodetics scientist of the highest order and
with vast experience. He was the last living associate of Gauß’ involved 
with the Hannover arc measurement and so completely
conversant with the method of his great mentor, who himself left no
documentation of it, that he might have shed some light upon questions that now
perhaps may forever remain in the dark}'' \cite{Baeyer1864}.

\section{The triangulation of Kurhessen}
\label{sec_triangulation_kurhessen}
\begin{figure}[h]
\begin{center}
\resizebox{0.75\columnwidth}{!}{\includegraphics{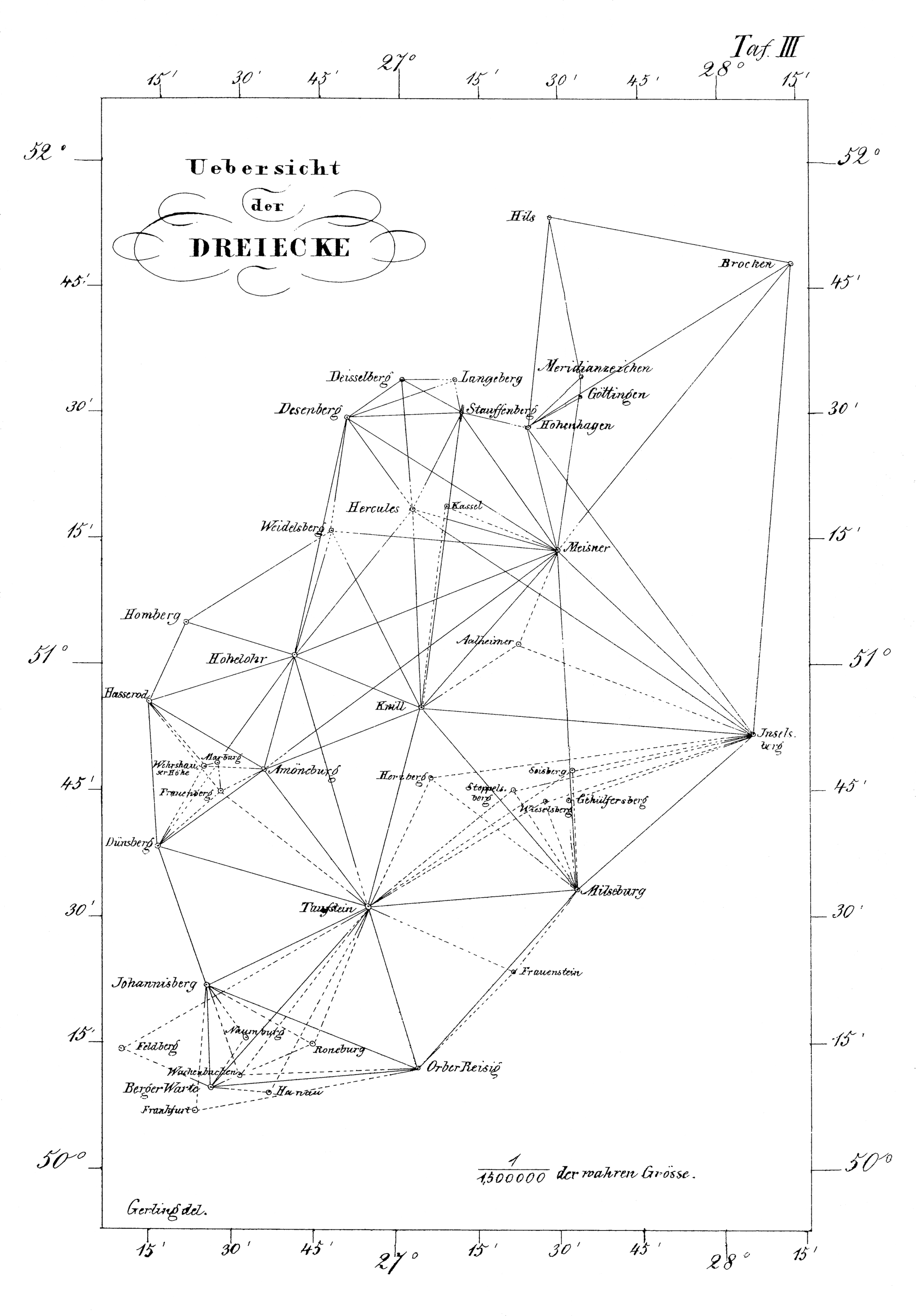}}
\caption{Network of the ''Kurhessische Triangulierung'' from 1822--1837
\cite{Gerling1839}}
\label{Fig_Kurhessische_Triangulation}
\end{center}
\end{figure}

In the spring of 1821, William II, elector of Hesse, sought Gerling's expertise
about the triangulation and topographical map of Kurhessen - the
electorate of Hesse.
After a first field exploration in fall 1821 and spring
1822 Gerling received the order of the triangulation of Kurhessen
[Gerling 1839;Reinhertz 1901]%
, which was carried out in two periods
from 1822 to 1824 and 1835 to 1837 (Fig.\ \ref{Fig_Kurhessische_Triangulation}).
In the northern region he included the triangle Brocken--Hohenhagen--Inselsberg,
which Gauß had measured during his Hannover arc measurement. In the south
Gerling connected his triangulation network with some of the survey marks close
to Frankfurt, Hesse, of a former Bavarian triangulation and a former
triangulation of the Grand Duchy of Hesse \cite{Torge2009}.

\begin{sloppypar}
For the triangulation Gerling used a twelve inch repeater theodolite from
Reichenberg--Ertel (Munich) and a ten inch universal theodolite from Breithaupt
(Cassel, Hesse). For local centering he used a copy of a Toise du Pérou, which
he had bought from Fortin, Paris, in 1831. Both the Breithaupt 
theodolite and the Toise are exhibited today in the
scientific instrument collection of the Physics Department of the
Philipps--Universität Marburg. Gerling's triangulation comprised twenty four
main geodetic reference points (first class) and seventeen marks of lower
precision (second class).
The base he used was the distance of Gauß's observatory to its meridian mark.
Gauß had deduced this distance from the base that Heinrich Christian Schumacher
had determined during the Holstein triangulation \cite{Torge2009}. For reducing 
the data Gerling used the reference ellipsoid of
Walbeck and the positional data of Göttingen, which Gauß had determined before.
He manually adjusted the network of the twenty four main geodetic
reference points (first class) in one procedure, a calculation of enormous
effort which Gauß gave him credit for (letter no.\ 290, \cite{Schaefer1927}).
The mean error of all direction in the triangulation network was $\pm
0\farcs88$, a recalculation for the ''Mitteleuropäische Gradmessung'' (Central
Eropean Arc Measurement) by Börsch gave an error of $\pm 0\farcs946$
\cite{Baeyer1866}.
\end{sloppypar}

Gerling's reduction is the first calculation of a triangulation network of Hesse
using an ellipsoid of rotation as reference. One hundred and seventy five years
after Gerling's pioneering work the ''Hessisches Landesamt für Bodenmanagement
and Geoinformation (HLBG)'' (Hessian State Authority of Real Estate Management
and Geoinformation) organized a regional survey of the preserved survey marks 
from Gerling's triangulation \cite{Heckmann2012}. Many of Gerling's marks 
were made of sandstone of considerable size and weight, therefore very steady.
The result of the recent survey is: fourteen of Gerling's marks of first--class
points and six marks of second--class points are still at their original
positions and are included in official evidence of geodetic reference points of
State of Hesse. Two further first--class marks and one second--class 
mark could be identified in the field and one other first--class point could 
be reconstructed exactly. A comparison of the marks' positions
in Gerling's triangulation with the high precision positions of the reference
system used currently has revealed a difference of less then 20 cm for most of
the marks and only in the rare case, at the edges of the network, more than 30 cm
\cite{Heckmann2012}.

\section{First evidence of the deflection of the vertical}
\label{sec_deflection_vertical}
The arc measurements of the 18th and the beginning of the 19th century gradually
revealed local deviations of an ellipsoid of rotation as the figure of the earth. To
precisely determine the moon's position, one needed to know the exact
local curvature. In 1810 Johann Georg von Soldner therefore suggested building
an observatory in Africa close to the equator for moon observations and 
conducting a precise arc measurement \cite{Torge2009}.
Pierre--Simon Laplace and von Soldner introduced the idea of a flattening which
must be described as a function of latitude and longitude
\cite{Torge2009}.
However, the known facts about the curvature of the earth at the beginning of
the 19th century were simply additional results from measurements of the
symmetric rotational figure of the earth.

In the 18th century the seconds pendulum became popular in the search for a
new measure of length. To a close approximation a pendulum of 1 m length has a
half period of oscillation of 1 second. The physical reason is that the
oscillation frequency depends on the gravitational acceleration and the mean
value at the surface of the earth is responsible for this finding. However, in
the early 19th century, scientists began to realize that a deviation from a
perfectly symmetric form of the earth might manifest itself not only in
differing curvature, but also differing gravitational acceleration. In 1818
Henry Kater explained for the first time the use of a modified seconds pendulum
for measuring the gravitational acceleration \cite{Kater1818}. However, this
pendulum was not easy to use in the fields.
The first movable instruments were developed in the 1860s, motivating more
gravitational measurements, but the breakthrough came in the 20th century, when
the free-fall gravimeters became available.

Most probably in 1827 C.F.\ Gauß undertook the first measurement of the
deflection of the vertical, which he explained in his report on determining
the latitude difference between the observatories in Göttingen and Altona
\cite{Gauss1828}. Using a Ramsden zenith sector he measured the zenith distance
of particular stars in both observatories. He then compared the astronomically
determined latitudes with those reduced from the triangulation of the kingdom of
Hannover from 1821--1824. The astronomical latitude difference was less by
5\farcs52. Also, with the pole height of the mountain Brocken (the highest peak
of the Harz mountain range), which had been astronomically determined by
Franz Xaver von Zach, he found a 10 -- 11\arcsec\ larger astronomical latitude
difference between Göttingen and the Brocken, and finally a 16\arcsec\ larger
astronomical latitude difference between Altona and the Brocken (for further
details see \cite{Wittmann2010}.) Gauß stated in his article, that this
difference did not seem very unusual; on the contrary, he expected such
differences to be found everywhere on the earth if the methods used to determine
them would be one or two orders of magnitude more precise. However, he continued
''\textit{that not until some time in the future centuries will the
mathematical knowledge of the figure of the earth be significantly advanced}''
%
\cite{Gauss1828}.

Unfortunately Gauß was right with this statement. It is striking that until the
beginning of the 19th century all reported astronomic--geodetic measurements
focused on the latitudes; all measurements of comparison were taken along the
same meridian, in the north--south direction. However, to correctly describe the
deflection of the vertical, one must also consider the deflection in longitude,
the east--west direction. In the second half of the 19th century Friedrich
Robert Helmert established the following definition of the deflection of the
vertical \cite{Torge2001}:
\begin{equation}
\label{eq:deflection_vertical}
\xi = \phi - \varphi \mbox{\hspace*{1cm}}  \eta = (\Lambda - \lambda)
\cos\phi
\end{equation}
with the astronomical latitude $\phi$ , the geodetic latitude
$\varphi$, the astronomical longitude $\Lambda$ and the geodetic
longitude $\lambda$.

For latitude measurements one had to determine zenith angles of stars, but for
measuring longitude differences the transit time of stars at the local meridian
had to be accurately observed. The precision of the longitude measurements
depends on the precision of the transit time measured. To keep the
error smaller than 0\farcs1, the time had to be determined accurately to
$0,1$/$15$ sec = $0,006$ seconds. In addition, not only the time measurement
itself, but also the time differences between transits at different locations on
the earth had to be measured with the same precision to determine a deflection
of the vertical in the range of 0\farcs1. Not an easy task in the 19th century.

In 1824 Friedrich Bernhard Nicolai organized longitude measurements in the area
of Mannheim \cite{Nicolai1825}, which can be regarded as precursor
experiments to Gerling's later measurements. Participating in a French
longitude campaign, Nicolai, together with the French colonel Henry, Johann
Gottlieb Friedrich von Bohnenberger and Friedrich Magnus Schwerd determined the
longitude differences between Straßbourg, Tübingen, Speyer and Mannheim by
synchronizing the observatory clocks via explosive signals --- a technique that
had been suggested before by von Zach and tested by Karl von Müffling
\cite{Berghaus1826} --- and by observing the transits of the same stars in all
of the four observatories. The local sidereal time was calculated from transits
of Bessel's fundamental stars. The differences between the transit times
directly gave the longitude differences. Comparing these with results from
triangulations (geodetic data), Nicolai could find an ''\textit{excellent
agreement}'' \cite{Nicolai1825} of geodetic and astronomic data for the
difference Mannheim--Straßbourg (geodetic 2\arcmin\ 54\farcs05), a minor
deviation of 0\farcs35 for the difference between Mannheim and Tübingen
(geodetic 2\arcmin\ 21\farcs91) and a deviation of 0\farcs16 for the difference
between Mannheim and Speier (geodetic 4\farcs90). Obviously Nicolai used the
astronomical measurements \textit{to confirm} the geodetic data; he did not
expect to observe a difference.

\section{Gerling's measurements at the Frauenberg in the summer of 1837}
\label{sec_gerlings_measurements_1837}

To complete the triangulation of Kurhessen Gerling decided in 1837 to organize an
astronomical longitude measurement across the entire network he had just
established \cite{Gerling1838}. As Nicolai before, Gerling was searching for a
control measurement. The larger and steadier instruments in the observatories
were expected to deliver more precise results. Therefore Gerling chose Göttingen
in the north east of his network and Mannheim in the south, approximately on the
same longitude as the Feldberg (Taunus) at the western edge. 
In Göttingen C.F.\ Gauß and his assistant
Carl Benjamin Goldschmidt could be convinced to participate, and in Mannheim
Gerling's colleague and friend Nicolai supported the campaign. In 1837
Gerling did not yet have an observatory in Marburg, therefore he chose the
Frauenberg, a small hill six km south--east of Marburg, as his temporary
observation site. A couple of years before, Gerling had used a flag signal at 
the ruin on
Frauenberg as survey mark of second class for the Kurhessian triangulation. In
1837 Gerling set a great stone post on top of the hill as a steady mount for his
theodolite. Furthermore, he transported his new high precision Box chronometer
from Kessels (Hamburg), which he just had bought for his planed observatory, to
the Frauenberg and raised a tent over the site. The task was to synchronize the
three observatory clocks and then to perform a series of transit observations of
the same stars in each of the participating observatories. These measurements
were scheduled for late summer in 1837, beginning on 24th August and ending
on 9th September.

In the 19th century observatories where very familiar with transit measurements,
or meridional measurements; this was the easy part of Gerling's campaign. Gauß
and Nicolai conveyed the sidereal times of their observatories to Gerling. At
the Frauenberg the stone post had unfortunately been set in soft ground and
Gerling detected that it was still moving during the campaign. He therefore used
corresponding solar altitudes to determine the local solar time, and from that
he calculated the sidereal time at Frauenberg.

The real challenge of the campaign was synchronizing the clocks to a
precision that would allow Gerling to detect a deviation between the 
geodetic and the astronomical longitude differences. The only way to
synchronize distant clocks in the first half of the 19th century was with light
signals. Because the Hohe Meißner in the northern part of Hesse could be seen
from Göttingen and Marburg, and the Feldberg was visible in Marburg and in
Mannheim, both mountains were used as signal stations. In the late afternoon a
co--worker at each of the two stations had to send heliotrope signals every 
8 minutes into both directions with an offset of 4 minutes in between the two 
stations. After nightfall a series of explosive signals were sent. This was
repeated on each day during the campaign in the late summer of 1837, 
whenever the weather conditions would allow measurements. 
During the day, corresponding solar
altitudes were recorded and at night the transits of stars observed; thus the
shift of the clocks could be monitored. Altogether 216 signals were sent from
the Meißner and 136 signals from the Feldberg station, of which 116 corresponding
signals from both stations finally formed the data base for synchronizing the
clocks.

Time measurements at that time were typically performed in the following way:
The observer was expecting a certain event, for example the transit of a star
passing a mark in the eyepiece of the telescope. When he recognized
the event, he notified a coworker who recorded the time from the
clock. There were two drawbacks to this method: First, because such a
measurement could not be double--checked, there was no way to assure accuracy.
Therefore measurements were repeated often in the hope that no or only
very few systematical errors would occur. Second, one systematic error could not
be excluded in this type of measurements: the reaction time of the observer, the
so--called ''personal equation''. Gerling visited his colleagues in Göttingen
and Mannheim and determined the personal equation of each observer by 
comparing meridional
observations of stars and by noting the time of passages of a pendulum. He could
thus estimate not only the statistical error, but also the offset due to the
reaction time. The offset turned out to be surprisingly large, which led Gerling
to mention it in a letter to Gauß (letter no.\ 294 \cite{Schaefer1927}) and to
question the former longitude determination of the island ''Helgoland'',
Greenwich and Paris.

After reducing the data Gerling published the following astronomically
determined longitude differences in units of time (24h = 360\degr)
\cite{Gerling1838}
\begin{tabbing}
Göttingen--Frauenberg: \hspace*{3em}	\= 4\min\ 36\fs19 $\pm$ 0\fs0152\\
Frauenberg--Mannheim:  					\> 1\min\ 19\fs67 $\pm$ 0\fs0208\\
Göttingen--Mannheim:					\> 5\min\ 55\fs86 $\pm$ 0\fs0258.
\end{tabbing}
The precision of the results is remarkable. Gerling achieved an error not
greater than 0.025 seconds on the time scale, which results in 0\farcs4 in
angular units.

\begin{figure}[h]
\centering
\resizebox{0.75\columnwidth}{!}
{\includegraphics{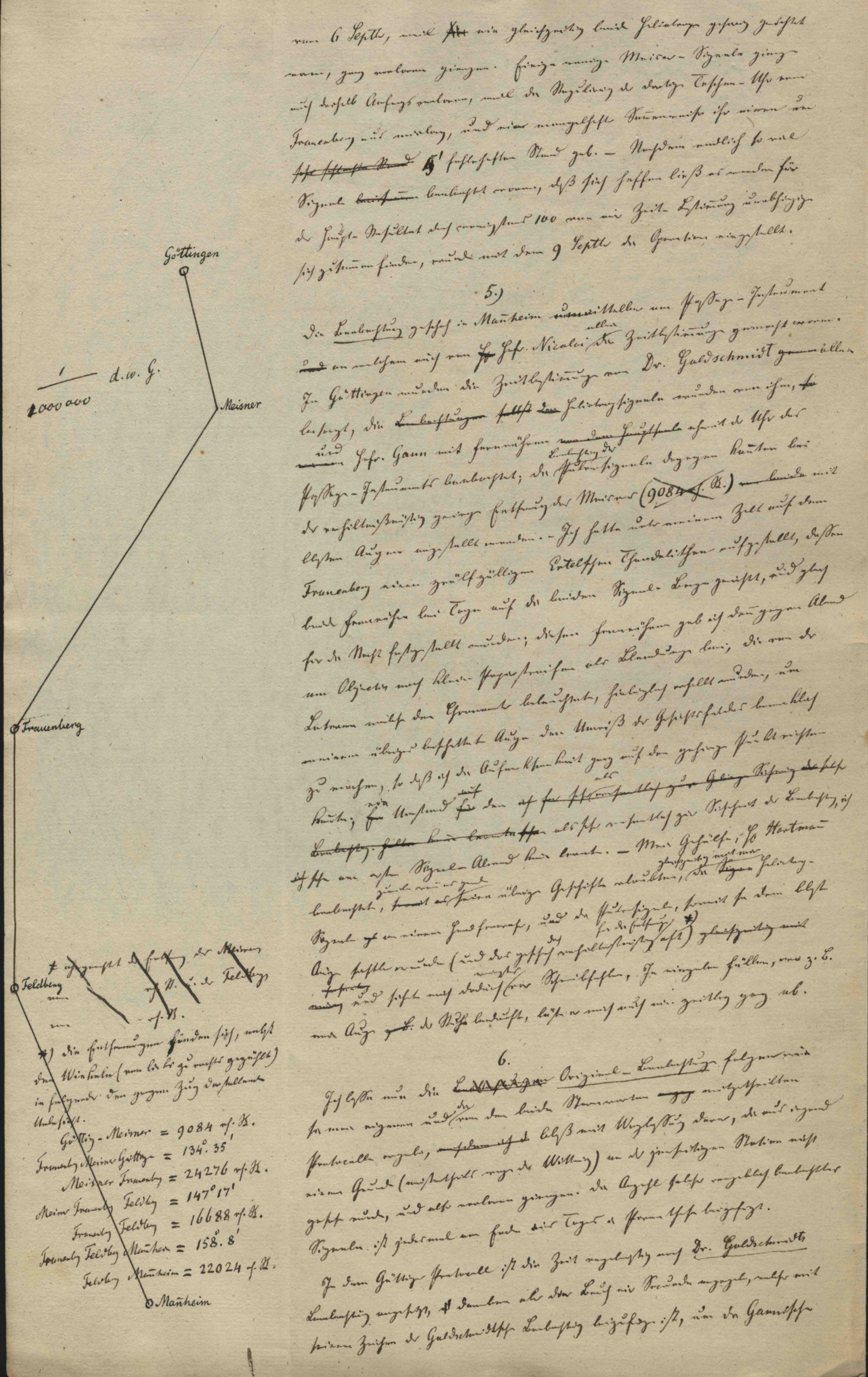}}
\caption{One page of the manuscript of Gerling's manuscript about his
measurements in 1837 \cite{Gerling1838} showing th beginning of chapter 5.
On the left a drawing of the stations involved in the campaign is included,
which he did not put in the article. (Archive of Chr.L.\ Gerling, Library of the
Philipps--Universität Marburg, sig.\ Ms.\ 352, page 3v.)}
\label{Fig_longitude_measurements_1837}
\end{figure}

With this result Gerling ended his article, which he submitted to the
\textit{Astronomische Nachrichten}. The significance of these measurements did
not become clear until the following contact with Gauß. In a letter dated 8
October 1838 (letter no.\ 294, \cite{Schaefer1927}), he mentioned noticeable
deviations of different longitude measurements that he used to confirm 
his results. Gauß pointed out that some of the measurements Gerling was citing
were geodetic and others astronomic longitude determinations and that he, Gauß,
did not expect them to show the same results. He mentioned his former latitude
measurements in northern Germany (letter no.\ 296, \cite{Schaefer1927}).
It was this remark of Gauß's, that opened Gerling's eyes: \textit{he had
performed the first measurement of the deflection of the vertical in longitude,
and the deviation of the astronomic and geodetic measurements was a new and very
valuable result}. His measurements initiated new quality of
measurements. He replied to Gauß (letter no.\ 297, \cite{Schaefer1927}):
''\textit{If I must accuse myself here of gross error and lack of thoroughness
in applying your § \em [symbol refers to a section of Gauß’ article from
1828]\em, then perhaps I can find comfort in that there are probably ''not
five persons existing in Europe'' who have taken heed of the § in this sense.  
\ldots I therefore feel compelled in this context to implement my own study in
another and more rational manner than I had originally intended with
insufficiently defined terms; and this is a great new merit which belongs to you
in this study.}''

%

In the final publication of the triangulation of Kurhessen
Gerling presented the longitude measurements of summer 1837 (\cite{Gerling1839},
page 204 ff.) in very different words. This time he indicated the difference
between astronomically and geodetically determined longitudes and listed the
deviations he found (Table \ref{Tab_longitude_deviations}). The deviations are
significant and are within the range of the deviations Gauß determined for
latitudes in northern Germany.

\begin{table}[h]
\centering
\caption{Astronomic and geodetic longitude differences Gerling determined in
summer 1837. In his report \cite{Gerling1839} Gerling did not include the
longitude differences between Frauenberg and Mannheim, but their deviation.
These data have been added for completeness.}
\label{Tab_longitude_deviations}
\begin{tabular}{llll}
\hline\noalign{\smallskip}
 & astronomic longitude & geodetic longitude & deviation \\
 & difference & difference & \\ 
 \noalign{\smallskip}\hline\noalign{\smallskip}
 
 Göttingen--Frauenberg & 1\degr\ 09\arcmin\ 02\farcs85 &	
 						 1\degr\ 09\arcmin\ 19\farcs49 & $- 16\farcs6$\\ 
 \noalign{\smallskip}\hline\noalign{\smallskip}						 
 Frauenberg--Mannheim &	\hspace{1em} 19\arcmin\ 55\farcs05 &
 					    \hspace{1em} 19\arcmin\ 42\farcs85 & $+ 12\farcs2$\\ 
\noalign{\smallskip}\hline\noalign{\smallskip}
 Göttingen--Mannheim  & 1\degr\ 28\arcmin\ 57\farcs90 &	
 					    1\degr\ 29\arcmin\ 02\farcs32 & $- 4\farcs4$\\ 
\noalign{\smallskip}\hline					    
\end{tabular}
\end{table}

\section{Comparison with later measurements}
\label{sec_comparison}

In 1841 Gerling could put his new observatory at the castle hill of Marburg
into operation. Via a local triangulation he determined the geodetic position of
the post on which the observatory's instrument was mounted \cite{Gerling1843b}:
longitude 26\degr\ 26\arcmin\ 2\farcs1 east of Ferro (Ferro was
used as reference meridian until 1884; the longitude of Ferro is 17\degr\
40\arcmin\ 00\arcsec\ west of Greenwich) and latitude 50\degr\ 48\arcmin\
46\farcs9 N.

The astronomical longitude of Gerling's observatory was determined by Ernst
Wilhelm Klinkerfues, a student of his who later became director of
the observatory in Göttingen and Gauß's successor. Klinkerfues reduced
observations of occultations of stars by the moon, which had been recorded and
published frequently and which were a valuable tool for calculating astronomical
longitudes. The result for the astronomical longitude of Gerling's observatory
was 18\min\ 28\farcs38 west of Berlin \cite{Gerling1855}. In this note
Klinkerfues was quoted: 
''\textit{However, because there are only very few locations where the longitude
is determined as well or even more accurately than that in Marburg, it seemed
ineffective to me, at least for the purpose I had restricted myself to, to
consider all observations. Even the corresponding observations made at major
astronomical observatories I did not include if, as in two cases, the
corrections to the tables from the Greenwich meridian observations were
known}''.
%
%
Again, the difficulties of high quality longitude determination is accentuated.

The astronomical latitude of Gerling's observatory was precisely determined
in 1862 by Richard Mauritius, one of his last doctoral students. Mauritius
used Bessel's method of measuring stars in the prime vertical, which
results in high precision pole height determination, and found the astronomical 
latitude to 50\degr\ 48\arcmin\ 44\farcs09. Using Klinkerfues' results and the
known geodetic position of the observatory in Berlin (30\degr\ 03\arcmin\
30\arcsec\ east of Ferro), he calculated a deviation of the longitude 
difference to Berlin of $+22\farcs2$ and a deflection of the vertical
in the latitude of $+2\farcs81$ \cite{Mauritius1862}.

The deviations of longitude and latitude differences determined by Gerling and 
his students must not be mistaken for deflection of the vertical data according
to the definition in equation (\ref{eq:deflection_vertical}). All data presented 
here so far are differences of two \textit{distant} stations; these
are not local deviations.
%
To collect accurate
deflection data of a certain location, the entire country, or
even better the entire planet, had to surveyed with a dense grid of
measuring points. For each point the
geodetic and astronomic position had to be measured and then a solution for all
points had to be calculated. This was an important aspect of the Central
European Arc Measurement organized by Johann Jacob Baeyer starting in
1862 \cite{Baeyer1861}. Unlike Gerling the scientists now could use telegraphic
signals for synchronizing their clocks, an enormous advancement, but also an
indication that Gerling's measurements were unique. However, the method of
observation did not change; they still had to deal with the reaction time of the
observers.
Friedrich Wilhelm Argelander, one of the participants and advisers of the
campaign very clearly specified that
%
''\textit{all pole heights and longitudes across the entire area of the arc
measurement are to be determined by the same observers, approximately four in
number, and with completely identical instruments}'', which unfortunately
''\textit{could not be conducted with absolute discipline}''
\cite{Hilfiker1885}.
Albrecht, Bruns and Hilfiker reduced the data of the Central European Arc
Measurement and published the astronomical longitudes of many of the European
observatories \cite{Hilfiker1885}. However, the net was too sparse, and the
number of points which should have been measured was too high for those
methods. It was not until the 20th century that scientists, with modern gravimeters
and zenith cameras, succeed in completing a dense network of gravimetric and
astronomic measurements and calculating maps of the vertical deflection with
sufficient precision. Gauß's prediction published in 1828 was
fulfilled in the 20th century.

In Table \ref{Tab_longitude_latitude_results} Gerling's results are compiled
together with results from the Central European Arc Measurement and modern
data. For comparison the astronomical data from Gerling and corresponding modern
data have been marked. The table shows the progress in methods of
data acquisition. The deviations of the positional data are greater than the
precision of the data, which suggests hidden systematic errors.

\begin{table}[h]
\rotatebox{90}{
\begin{minipage}{\textheight}
\centering
\caption{Survey of different latitude and longitude results for
Göttingen, Marburg and Mannheim. The geodetic coordinates on ellipsoid GRS80
are related to the European Terrestrial Reference System 1989 (ETRA89). These 
coordinates were provided by
the Hessisches Landesamt für Bodenmanagement und Geoinformationen (HLBG) and the
Landesamt für Geoinformationen und Landentwicklung Niedersachen (LGLN), as well
as the Landesamt für Geoinformationen und Landentwicklung Baden--Würtemberg
(LGL). The data of the vertical deflection with respect to GRS80 were
calculated in consideration of the local mass distribution in the surroundings
of the listed locations by the Bundesamt für Kartographie und Geodäsie (BKG). }
\label{Tab_longitude_latitude_results}
\renewcommand{\arraystretch}{1.15}
\begin{tabular}{c|cccc|cc}
\hline
 & Göttingen (GÖ) & Frauenberg (FR) & Marburg (MR) & Mannheim (MN) & 
 \multicolumn{2}{c}{longitude difference} \\ 
 & observatory & measuring post & observatory & observatory & 
 GÖ--MN & GÖ--FR \\ 
 \hline
 geodetic latitude & 51\degr\ 31\arcmin\ 47\farcs850 & 50\degr\ 45\arcmin\
 27\farcs751 & 50\degr\ 48\arcmin\ 46\farcs884 & 49\degr\ 29\arcmin\
 14\farcs681 & & \\
 Walbeck & & & & & & \\
 \hline
 %
astronomic latitude & & &\cellcolor[gray]{0.9} 50\degr\ 48\arcmin\
44\farcs09\phantom{0} & & &\\
1862 & & &\cellcolor[gray]{0.9}  \cite{Mauritius1862} & & & \\
\hline
%
geodetic latitude & 51\degr\ 31\arcmin\ 42\farcs943 & 50\degr\ 45\arcmin\
23\farcs494 & 50\degr\ 48\arcmin\ 42\farcs583 & 49\degr\ 29\arcmin\
11\farcs385 & & \\
GRS80 & & & & & & \\
\hline
 %
deflection of the & $+4\farcs813$ & $-0\farcs168$ &
$+0\farcs765$ & $-0\farcs013$ & & \\ 
vertical $\xi$, BKG 2012 & & & & & & \\
\hline
 %
astronomic latitude & 51\degr\ 31\arcmin\ 47\farcs756 & 50\degr\ 45\arcmin\
23\farcs326 & \cellcolor[gray]{0.9} 50\degr\ 48\arcmin\ 43\farcs348 & 49\degr\
29\arcmin\ 11\farcs372 & & \\  today & & & \cellcolor[gray]{0.9} & & & \\
\hline\hline
%
geodetic longitude & 27\degr\ 36\arcmin\ 28\farcs200 & 26\degr\ 27\arcmin\
08\farcs712 & 26\degr\ 26\arcmin\ 02\farcs100 & 26\degr\ 07\arcmin\
27\farcs712 & 1\degr\ 29\arcmin\ 02\farcs32\phantom{0} & 1\degr\ 09\arcmin\
19\farcs49\phantom{0} \\
Walbeck & & & & & & \\
\hline
 %
astronomic longitude & 27\degr\ 36\arcmin\ 26\farcs40\phantom{0} & 26\degr\
27\arcmin\ 23\farcs55\phantom{0} & 26\degr\ 26\arcmin\ 24\farcs3\phantom{00} &
26\degr\ 07\arcmin\ 28\farcs5\phantom{00} &
\cellcolor[gray]{0.9} 1\degr\ 28\arcmin\ 57\farcs90\phantom{0} &
\cellcolor[gray]{0.9} 1\degr\ 09\arcmin\ 02\farcs85\phantom{0} \\
1839/1862 & (ref.\ Mannheim) & (ref.\
Mannheim) & (1862, ref.\ Berlin) & (Nicolai/Wurm) & \cellcolor[gray]{0.9} &
\cellcolor[gray]{0.9}\\
\hline
%
astronomic longitude & 27\degr\ 36\arcmin\ 35\farcs944 & & & 26\degr\ 07\arcmin\
38\farcs689 & 1\degr\ 28\arcmin\ 58\farcs255 & \\
1885 & & & & & & \\
\hline
%
geodetic longitude & 27\degr\ 36\arcmin\ 34\farcs022 & 26\degr\ 27\arcmin\
15\farcs677 & 26\degr\ 26\arcmin\ 09\farcs080 & 26\degr\ 07\arcmin\
34\farcs912 & 1\degr\ 28\arcmin\ 59\farcs110 & \\
(Ferro) GRS80 & & & & & & \\
\hline
%
deflection of the vertical & $-0\farcs775$ & $+4\farcs651$ & $+6\farcs566$ 
& $+1\farcs184$ & & \\ 
$\eta/\cos\varphi$, BKG 2012 & & & & & & \\
\hline
%
astronomic longitude & 27\degr\ 36\arcmin\ 33\farcs247 & 26\degr\ 27\arcmin\
20\farcs328 & 26\degr\ 26\arcmin\ 15\farcs646 & 26\degr\ 07\arcmin\
36\farcs096 & \cellcolor[gray]{0.9} 1\degr\ 28\arcmin\ 57\farcs151 
& \cellcolor[gray]{0.9} 1\degr\ 09\arcmin\ 12\farcs919 \\
(Ferro) today & & & & &\cellcolor[gray]{0.9} &\cellcolor[gray]{0.9} \\
\hline
\end{tabular}
\end{minipage}
}
\end{table}

\section{Conclusion}
\label{sec_conclusion}

\begin{sloppypar}
Gerling determined the astronomic longitude difference between Göttingen and
Mannheim with a small deviation of 0\farcs75 compared to modern data, which is
remarkable considering the methods he used. In contrast, the longitude
difference between Göttingen and Gerling's station at Frauenberg shows a noticeable
deviation of 10\farcs07. Gerling calculated the difference between Göttingen and
Mannheim as the sum of the differences Göttingen--Frauenberg and
Frauenberg--Mannheim. Therefore the larger deviations to Frauenberg, which
cancel out in the sum, most probably reveal a systematic error in the local
sidereal time or the mean solar time at Frauenberg of 0\fs67. In his article in
the \textit{Astronomische Nachrichten} Gerling mentioned the instability of the
post he placed at the Frauenberg for his theodolite. Instead of
meridional observations of stars he used corresponding solar altitudes to
determine the mean solar time. He measured the height of the sun with a 
prism sextant and an artificial horizon, and calculated the mean solar
time to control the shift of his clock. He was able to detect a jump in the
shift of 0.2 sec on a cold and windy day. However, it seems reasonable that
using a small instrument to determine the local time will not result in
the same precision achievable with a large instrument on a steady post in an 
observatory. The sidereal times of Göttingen and Mannheim were
very accurate and served as a perfect base for Gerling's results.
\end{sloppypar}

\begin{sloppypar}
Gerling's measurements of the astronomical longitude difference between
Göttin\-gen and Mannheim were of unprecedented precision. Synchronizing the
clocks proved to be a worthwhile effort. Contrary to Gauß's opinion, Gerling could
demonstrate that even with the methods available in the first half of the 19th
century the deflection of the vertical on both latitude and longitude could be
determined. In this regard Gerling deserves to be honored alongside C.F.\ Gauß
in the history of progress to precisely determine the figure of the earth!
\end{sloppypar}

\begin{acknowledgement}
\textit{Acknowledgements}. 
I appreciate the cooperation with Bernhard Heckmann, Hessisches Landesamt für
Bodenmanagement und Geoinformationen, Wiesbaden. In discussions he gave valuable
hints and corrections and he obtained the modern positional and deflection of
the vertical data. Also, I'd like to thank the Bundesamt für Kartographie und
Geodäsie (BKG), Außenstelle Leipzig for calculation and provision of recent
deflection of the vertical data. Thanks to Judith Whittaker--Stemmler for
translating the quotations of the original German articles.
\end{acknowledgement}

\end{document}